\documentstyle[preprint,aps]{revtex}
\newcommand{\be}{\begin{equation}} 
\newcommand{\ee}{\end{equation}} 
\newcommand{\ben}{\begin{eqnarray}} 
\newcommand{\een}{\end{eqnarray}} 
\begin{document} 

\tighten

\begin{center} 
{\bf Topological solitons in a vacuumless system}\footnote{This 
work is partially supported by the U. S. Department of Energy (D.O.E.) 
under cooperative research agreement DE-FC02-94ER40818, and by Conselho
Nacional de Desenvolvimento Cient\'\i fico e Tecnol\'ogico, CNPq, Brazil.} 
\end{center} 
 
\begin{center} 
D. Bazeia\\
Center for Theoretical Physics\\ 
Laboratory for Nuclear Science and Department of Physics\\ 
Massachusetts Institute of Technology, Cambridge, Massachusetts 02139-4307\\
and\\
Departamento de F\'\i sica, Universidade Federal da Para\'\i ba\\
Caixa Postal 5008, 58051-970, Jo\~ao Pessoa, Para\'\i ba, Brazil
\end{center} 
 
\begin{abstract}
 
We investigate a model for a real scalar field in bidimensional 
space-time, described in terms of a positive semi-definite potential that 
presents no vacuum state. The system presents topological solutions
of the BPS type, with energy density that follows a Lorentzian law.
These BPS solutions differ from the standard $\tanh$-type kink,
but they also support bosonic and fermionic zero modes.
\end{abstract}

\vskip 1.5cm

\begin{center} 
\end{center} 
\maketitle

\newpage 
 
Topological defects such as domain walls, strings, monopoles, 
can appear in models of symmetry-breaking phase transitions in the early 
universe \cite{vsh94}. These standard defects appear in models 
where the potential presents at least two degenerate vacuum states. For 
instance, domain walls require a countable set of vacuum states, in models 
where one breaks some discrete symmetry, and cosmic strings and monopoles 
require uncountable set of vacuum states, in models where 
one breaks some continuum symmetry. 
 
Despite this standard situation, however,        
there are other models such as the ones considered for instance in 
Refs.~{\cite{dja82,cvi98}}, which engender interesting features.
For instance, in \cite{dja82} one considers 
the Liouville theory to offer a way of breaking spontaneously the tranlation 
invariance of the spatial portion of space-time.  More recently, in 
\cite{cvi98} one considers other forms of systems defined by vacuumless 
potentials. In the present work we shall be mainly concerned with 
the model defined by 
\be 
\label{L} 
{\cal L}=\frac{1}{2}\,\partial_{\alpha}\phi\,\partial^{\alpha}\phi-U(\phi) 
\ee 
with the potential 
\be 
\label{model} 
U(\phi)=\frac{1}{2}\,\frac{\mu^2}{\lambda^2}\,\,{\rm sech}^2(\lambda\,\phi) 
\ee 
We consider $\mu$ real and positive, and $\lambda$ real. The potential
is positive semi-definite, presents a maximum at $\phi=0$ and no vacuum state. 
Despite the absence of vacuum states, the system is still able to support 
topological defects. This model was recently investigated in \cite{cvi98}, 
and also in \cite{cvi98a}, and there attention was given mainly on issues 
concerning gravitational aspects of the new topological defect. 
In the present work, however, we shall focus attention on exposing other 
features of the system, considering the $1+1$ dimensional Minkowsky
space-time: $x^0=x_0=t$ and $x^1=-x_1=x$. Here $\lambda$ is dimensionless,
and $\mu$ has dimension inverse of distance, dim($\mu$)=1/dim($x$). The
field $\phi$ is dimensionless, and the system behaves standardly in
bidimensional space-time. 
 
The equation of motion is 
\be 
\label{soeqt} 
\frac{\partial^2\phi}{\partial t^2}-\frac{\partial^2\phi}{\partial x^2}- 
\frac{1}{\lambda}\,\mu^2\,{\rm sech}^2(\lambda\,\phi)\tanh(\lambda\,\phi)=0 
\ee 
and for static $\phi=\phi(x)$ we get 
\be 
\label{soeq} 
\frac{d^2\phi}{dx^2}=-\frac{1}{\lambda}\,\mu^2\,
{\rm sech}^2(\lambda\,\phi)\tanh(\lambda\,\phi) 
\ee 
As shown in Ref.~{\cite{cvi98}}, this equation is solved by 
\be 
\label{soliton} 
{\phi}(x)=\pm\,\frac{1}{\lambda}\,{\rm arcsinh}(\mu x) 
\ee 
These solutions diverge asymptotically, and their specific forms depend 
on $\mu$ and $\lambda$. Although they are somehow 
similar to the usual $\tanh$-type kink and antikink that appear is the 
$\phi^4$ model, they are much more diffuse than the standard 
$\tanh$-type defect, and have divergent amplitude.  

To expose key features of this classical solution, let us consider 
the energy of static configurations. Here we have 
\be 
\label{energy} 
E=\frac{1}{2}\int^{\infty}_{-\infty}dx\left(\,\left(\frac{d\phi}{dx}\right)^2+ 
\frac{\mu^2}{\lambda^2}\,{\rm sech}^2(\lambda\,\phi)\right) 
\ee 
We can write 
\be 
E=E_B+\frac{1}{2}\int^{\infty}_{-\infty}dx 
\left(\frac{d\phi}{dx}-\frac{\mu}{\lambda}\,{\rm sech}(\lambda\,\phi)\right)^2 
\ee 
$E_B$ is the value that minimizes the energy. It can be written as 
$E_B=\Delta W$, where 
\be 
\Delta W=W[\phi(x\to\infty)]-W[\phi(x\to-\infty)] 
\ee 
It only depends on the asymptotic values of the function 
\be 
W(\phi)=\frac{\mu}{\lambda^2}\arctan[\sinh(\lambda\,\phi)] 
\ee 
The procedure here is similar to the case of coupled fields, as 
considered in Refs.~{\cite{bsr95,bsa96} and in applications 
to condensed matter \cite{cond} and field theory itself \cite{field}. 
 
The function $W(\phi)$ obeys 
\be
\label{dw}
\frac{dW}{d\phi}=\frac{\mu}{\lambda}\,{\rm sech}(\lambda\,\phi) 
\ee 
and the potential in $(\ref{model})$ can be written as 
\be 
U(\phi)=\frac{1}{2}\left(\frac{dW}{d\phi}\right)^2 
\ee 
For static fields the energy is bounded to $E=E_B$ for field configurations 
that solve the first-order equation 
\be 
\label{foeq} 
\frac{d\phi}{dx}=\frac{\mu}{\lambda}\,{\rm sech}(\lambda\,\phi) 
\ee 
We see that solutions of this first-order equation also solve the 
equation of motion. Yet, we can check explicitly that 
\be
\label{cs} 
{\bar{\phi}}(x)=\frac{1}{\lambda}{\rm arcsinh}(\mu x) 
\ee 
solves the first-order equation $(\ref{foeq})$. 
This means that this solution is stable \cite{bsa96} 
and of the BPS \cite{bps75} type \cite{field,csh97}. We return to the issue
of stability below, making the argument explicit.
 
The BPS solutions present the interesting feature of allowing 
the energy to be written as
\be 
E_B=\int_{-\infty}^{\infty}dx\left(\frac{d{\bar\phi}}{dx}\right)^2= 
\frac{\mu^2}{\lambda^2}\,\int_{-\infty}^{\infty}\,dx\,
{\rm sech}^2({\lambda\,\bar\phi}) 
\ee 
The energy density has the form 
\be
\label{ed}
\varepsilon(x)=\frac{1/\lambda^2}{x^2+1/\mu^2} 
\ee 
Interestingly, the energy density of the BPS solution obeys a Lorentzian law, 
and accordingly is more diffuse than the standard $\tanh$-type 
kink. Explictly, the $\tanh$-type kink appears when the potential is
\be
U_s(\phi)=\frac{1}{2}\lambda^2(\phi^2-a^2)^2
\ee
Here $a$ is real and positive, and the kink solutions
are $\phi(x)=\pm\,a\,\tanh(\lambda a x)$. The corresponding
energy density reads
\be
\varepsilon_s(x)=\frac{1}{2}\,\lambda^2\,a^2\,{\rm sech}^4(\lambda\,a\,x)
\ee
In spite of this, however, the energy density $\varepsilon(x)$ in
(\ref{ed}) is still integrable and gives the finite value
$E_B=\mu\,\pi/\lambda^2$. Alternatively, although the classical static
solution ${\bar\phi}(x)$ diverges asymptotically, the asymptotic values of
$W(\phi)$ are finite and give a well-defined $\Delta W$. This fact can be used 
to expose the topological aspects of the solution. In $1+1$ dimensions we can 
introduce the topological current 
\be 
J^{\mu}=\epsilon^{\mu\nu}\partial_{\nu} W(\phi) 
\ee 
Here we are following the first work in Ref.~{\cite{cond}},
using $W(\phi)$ to define the current; it makes the conserved
(topological) charge identical to the energy.
This procedure of using $W(\phi)$ is in general more
appropriate than considering the field $\phi$ itself, as it
is usually done in the case of the standard $\tanh$-type solution.
This is evident here, since the classical solution diverges
asymptotically and would make the charge (artificially) ill-defined
if one uses $\epsilon^{\mu\nu}\partial_{\nu}\phi$ to define
the topological current.

To investigate classical or linear stability, or to compute 
the first quantum corrections \cite{jac77,raj82} introduced by the
bosonic field $(\ref{cs})$ we consider
\be
\label{flu}
\phi(x,t)={\bar{\phi}}(x)+\sum_n\,\eta_n(x)\,\cos(w_n t) 
\ee 
We substitute this into the time-dependent equation of motion
$(\ref{soeqt})$, and consider the case of small fluctuations $\eta_n(x)$
about the classical field ${\bar{\phi}}(x)$ to get 
\be
\label{se}
\left(-\frac{d^2}{dx^2}+V(x)\right)\,\eta_n(x)=w_n^2\,\eta_n(x) 
\ee
where
\be
V(x)=\frac{-1/\mu^2+2x^2}{(x^2+1/\mu^2)^2}
\ee
is the potential of this Schr\"odinger-like equation.

Stability of the classical solution $(\ref{cs})$ implies that $w_n$ in
Eq.~(\ref{flu}) should be real. This makes $w_n^2\ge0$, and so the
Schr\"odinger-like Hamiltonian that appears in Eq.~({\ref{se}}) has to be
positive semi-definite. But this is indeed the case, since we can factorize
the Hamiltonian in Eq.~(\ref{se}) in a very specific way. To show this
explicitly we use the equations $(\ref{dw})$ and $(\ref{foeq})$ to write
\be
\frac{d\phi}{dx}=\frac{dW}{d\phi}
\ee
This equation can be used to introduce the operators 
\ben 
\label{a} 
a_{\pm}&=&\pm\,\frac{d}{dx}+W_{\phi\phi}\nonumber 
\\&=& \pm\,\frac{d}{dx}-\mu\,\,
{\rm sech}({\lambda\,\phi})\,\tanh({\lambda\,\phi}) 
\een 
where $W_{\phi\phi}$ stands for $d^2W/d\phi^2$. These operators obey
$a_{\pm}^{\dagger}=a_{\mp}$, and can be used to introduce
$H_{+}=a_{+}^{\dagger}a_{+}$ and $H_{-}=a_{-}^{\dagger}a_{-}$
as the Hamiltonians 
\be 
\label{h} 
H_{\pm}=-\frac{d^2}{dx^2}+ V_{\pm} 
\ee 
where 
\be 
V_{\pm}=W^2_{\phi\phi}\mp W_{\phi}\,W_{\phi\phi\phi} 
\ee 
We use the classical solution $(\ref{cs})$ to write these potentials
in the explicit forms
\ben 
\label{+} 
V_{+}(x)&=& \frac{1/\mu^2}{(x^2+1/\mu^2)^2}
\\ 
\label{-}
V_{-}(x)&=& \frac{-1/\mu^2+2x^2}{(x^2+1/\mu^2)^2}
\een 
The potential $V_{-}(x)$ is exactly $V(x)$, the potential that appears
in the Schr\"odinger-like equation $(\ref{se})$. We use this 
result and $|n>=\eta_n(x)$ and $<n|m>=\delta_{nm}$ to write
\ben
w^2_n&=&<n|\left(-\frac{d^2}{dx^2}+
\frac{-1/\mu^2+2x^2}{(x^2+1/\mu^2)^2}\right)|n>\nonumber\\
&=&<n|\left(\frac{d}{dx}-\frac{x}{(x^2+1/\mu^2)}\right)
\left(-\frac{d}{dx}-\frac{x}{(x^2+1/\mu^2)}\right)|n>\nonumber\\
&=&\int^{\infty}_{-\infty}dx\,|\zeta_n(x)|^2
\een
where
\be
\zeta_n(x)=\left(-\frac{d}{dx}-\frac{x}{(x^2+1/\mu^2)}\right)\eta_n(x)
\ee
This result shows that $w^2_n$ can not be negative.
This proof follows the work in Ref.~{\cite{bsa96}}, where it was done
in general, for systems of coupled real scalar fields.

The Schr\"odinger-like equation $(\ref{se})$ has at least one bound state,
the bosonic zero mode that is present due to translational invariance.
The normalized eigenfunction $\eta_0(x)$ is given by 
\be 
\label{bzm} 
\eta_0(x)=\frac{\sqrt{(1/\mu\,\pi)\,}}{\sqrt{(x^2+1/\mu^2)\,}} 
\ee 
It is not hard to see that there is no other bound state. In this case 
no meson can be binded to the soliton, and only scaterring states
may appear \cite{jac77,raj82}.

The Lagrangian density $(\ref{L})$ can be seen as the bosonic portion 
of a supersymmetric theory, and in the extended supersymmetric model the 
function $W=W(\phi)$ plays the role of the superpotential -- see for 
instance Ref.~{\cite{csh97}}. We can introduce fermions with the 
usual Yukawa coupling $f(\phi)\,{\bar\psi}{\psi}$. We may consider 
$f(\phi)=g\,\phi$ or 
\be
\label{c} 
f(\phi)={g}\,\frac{d^2W}{d\phi^2}= 
-g\,\mu\,{\rm sech}(\lambda\,\phi)\,\tanh(\lambda\,\phi) 
\ee 
The possibility of introducing fermions in a sypersymmetric way requires 
the coupling $(\ref{c})$, with $g=\pm1$ and Majorana spinors.
In the model with fermions we search for fermionic zero modes. We consider
the more general case where the spinors are Dirac spinors.
Here the relevant Dirac equation is
\be
i\gamma^1\frac{d\psi}{dx}+f(\phi)\,\psi=0
\ee
We use $\psi_{\pm}$ as the upper $(+)$ 
and lower $(-)$ components of the Dirac spinor $\psi$, and the representation 
$i\gamma^1\to\sigma_3$ to get 
\be 
\pm \frac{d\psi_{\pm}}{dx}+f(\phi)\,\psi_{\pm}=0 
\ee 
For $f(\phi)$ given by Eq.~(\ref{c}), for instance, the fermionic zero mode 
depends on $\mu$ and $g$, and may not exist for specific values 
of $g$, as for instance for $|g|=1/2$. There are fermionic zero modes
for $|g|>1/2$. For $g>1/2$ we get
\be 
\psi_0=C(\mu,g)\,\sqrt{\left(\frac{1}{(x^2+1/\mu^2)} 
\right)^{g}}\,\left({0\atop 1}\right) 
\ee 
and for $g<1/2$ we get 
\be 
\psi_0=C(\mu,g)\,\sqrt{\left(\frac{1}{(x^2+1/\mu^2)} 
\right)^{-g}}\,\left({1\atop 0}\right) 
\ee 
where $C(\mu,g)$ is the normalization constant,
which depends on both $\mu$ and $g$. For instance,
$C(\mu,g=\pm1)= \sqrt{1/\mu\,\pi\,}$,
and in this case the fermionic zero mode very much resembles
the bosonic zero mode $(\ref{bzm})$: this reflects the fact that
$g=\pm1$ are the only two values of $g$
that allow the supersymmetric extension. 

\acknowledgments

The author would like to thank Roman Jackiw for discussions, and the 
Center for Theoretical Physics, MIT, for hospitality.

\end{document}